\documentclass[11pt,a4]{article}
\usepackage{epsfig}    
\usepackage{amsmath}   
\usepackage{amssymb}  
    
\newcommand{\beeq}{\begin{eqnarray}}    
\newcommand{\eeeq}{\end{eqnarray}}    
\newcommand{\be}{\begin{equation}}    
\newcommand{\ee}{\end{equation}}

\def\omp{\omega_{I\!\!P}}  
    
\def\xb{{\bf x}}    
\def\yb{{\bf y}}    
\def\zb{{\bf z}}   
   
\def\rb{{\bf r}}    
\def\bb{{\bf b}}    
    
\def\qbar{\bar{q}}    
\def\asb{\overline{\alpha}_s}

\newcommand{\hep}[1]{{\tt hep-ph/#1}}

\newcommand\npb[3]{{\it Nucl. Phys. }{\bf B #1} (#2) #3}   
\newcommand\npa[3]{{\it Nucl. Phys. }{\bf A #1} (#2) #3}   
   
\newcommand\plb[3]{{\it Phys. Lett. }{\bf B #1} (#2) #3}   
   
\newcommand\epjc[3]{{\it Eur. Phys. J.}{\bf C #1} (#2) #3}   
\newcommand\prd[3]{{\it Phys. Rev. }{\bf D #1} (#2) #3}   
   
\newcommand\zpc[3]{{\it Z. Phys. }{\bf C #1}(#2) #3}   
    
\begin{document}

\titlepage    
\begin{flushright}    
DESY 03--075 \\    
June 2003    
\end{flushright}    
    
\vspace*{1in}    
\begin{center}    
{\Large \bf On solutions of the Balitsky-Kovchegov equation with    
impact parameter}\\    
\vspace*{0.4in}    
K. \ Golec--Biernat$^{(a,b)}$    
and A.\ M. \ Sta\'sto$^{(c,a)}$ \\    
\vspace*{0.5cm}    
$^{(a)}${\it Institute of Nuclear Physics, Radzikowskiego 152,    
 Krak\'ow, Poland}\\   
{\it $^{(b)}$ II.\ Institut f\"ur Theoretische Physik,    
    Universit\"at Hamburg, Germany}\\    
$^{(c)}$ {\it DESY, Theory Division, Notkestrasse 85, 22603 Hamburg, Germany} \\    
 \vskip 2mm    
\end{center}    
\vspace*{1cm}    
\centerline{(\today)}    
    
\vskip1cm    
\begin{abstract}    
We analyze numerically the   
Balitsky-Kovchegov equation 
with the  full impact parameter dependence $b$.   
We show that due to the particular $b$-dependence of the initial condition    
the amplitude  decreases for large dipole sizes $r$.  
Thus the region of saturation 
has a finite extension in the dipole size $r$,  
and its width increases with rapidity. We calculate  
the $b$-dependent saturation scale and discuss limitations on geometric scaling. 
We also demonstrate the instant emergence of the power-like tail in impact parameter,    
which  is due to the  long range contributions.    
Thus the resulting cross section violates Froissart bound  despite    
the presence of  a nonlinear term responsible for saturation.

\end{abstract}

\newpage    
    
\section{Introduction}    
\label{sec:1}    
    
 The  high energy limit of QCD is one of the most intriguing aspects of 
hadronic physics. With the advent of new generation 
accelerators like 
HERA,  Tevatron, RHIC and in a near future LHC, the basic problems 
of strong interactions are experimentally studied and confronted 
with theoretical predictions of high energy QCD. The important 
discovery of the rise of the proton structure functions at HERA \cite{HERA}  
at small values of the Bjorken variable $x$ 
(which is equivalent to the high energy limit) is an example of such a 
confrontation. 
This rise was predicted by high energy QCD \cite{BFKL} and is related to 
the increase of  the gluon density. 
Ultimately the rise has to be damped by the presence of the saturation effects 
which enter via nonlinear modification to the QCD evolution as first
proposed in the pioneering work  \cite{GLR}. 
Thus QCD at high energy is the theory of high density systems of colored 
particles. 
In such systems hard scales appear which allow to apply perturbative 
techniques although 
effective interactions in the dense partonic system are of 
non-perturbative origin. 
The interplay between hard and soft (perturbative and non-perturbative) 
aspects of QCD,  which also touches in the operational way the issue of confinement, 
is the most exciting element of high energy QCD.

The effective theory which describes high energy scattering in QCD is 
Color Glass Condensate \cite{MCLERVEN}. The basic equations of this theory \cite{JAMAL} are equivalent \cite{MUELLER2001} 
to the hierarchy of equations derived by Balitsky \cite{BAL1} and later on reformulated 
in a compact form by Weigert \cite{WEIGERT}. These equations contain the BFKL evolution and also  
the  triple Pomeron vertex  \cite{BARWUE}. 
In this work we present the results of numerical studies of these 
basic equations. To be precise 
we study the simplified version of the hierarchy of Balitsky's equations 
which reduces to one equation 
in the limit of large number of colors. This is the equation obtained by 
Kovchegov \cite{KOV1} in the dipole 
approach \cite{Mueller:94} to high energy scattering in QCD. With this equation the deep 
inelastic lepton--nucleus 
scattering can be described and also information on small-$x$ hadronic 
wave function be obtained.

We study this equation in the full form, including the impact parameter 
dependence $b$. 
Previous analytical \cite{KOV2,LEVTUCH} and numerical \cite{BRAUN,LEVLUB, GBMS} studies 
were done under a simplified assumption of infinitely large and uniform 
nucleus, i.e. neglecting 
impact parameter $b$. Recently, an approximate
 solution to this equation in semiclassical 
approach was considered \cite{BoKoLe03}. As we will show, the solutions of the full form of the 
Balitsky-Kovchegov 
(BK) equation possess important new  features in comparison to the uniform 
case, e.g. 
restricted scaling properties with impact parameter dependent 
saturation scale. 
The detailed analysis of the $b-$dependence of the solution  allows to study the high energy  
behaviour of the $\gamma^*N$ cross section. We show 
that the Froissart bound \cite{FrMa} is violated due to the long range contributions 
in the kernel \cite{KW} 
which brings the issue of the lack 
of confinement effects in the BK equation. 
The problem of the Froissart bound was also extensively discussed in the
same context in \cite{KW,FIIM} based on analytical consideretations.

The paper is organised in the following way. In the next section we 
briefly present the BK equation 
and its symmetries. In Sec.~3 we describe the numerical methods of finding 
the solution and discuss the  initial condition. In Sec.~4 we present the resulting amplitude $N$ as  
a function of dipole size $r$ and extract the $b$ - dependent saturation scale.  
In Sec.~5 we discuss the form of the impact parameter profile 
which emerges in the evolution, and in particular we concentrate on the emergence  
of the power tails in $b$ in the amplitude. We also present the estimate of the cross section  
of the black disc radius and its dependence on the rapidity. 
Finally, in Sec.~6 we state our conclusions.

\section{The  Balitsky-Kovchegov equation}    
\label{sec:2}    
    
The deep inelastic scattering of a lepton on a nucleus at high energy    
in the dipole picture \cite{Mueller:94,NiZa:91} is viewed    
in the nucleus rest frame as the splitting of an exchanged virtual photon into    
a $q\qbar$ dipole and  the subsequent interaction of the dipole with the nucleus.    
The latter process is described by the dipole-nucleus scattering amplitude    
$N(\xb,\yb)$, where $\xb,\yb$ are two-dimensional vectors of  the transverse position   
of the dipole  ends. Alternatively, one can introduce the dipole vector    
$\rb=\xb-\yb$, and  the impact parameter $\bb=(\xb+\yb)/2$.    
Thus in general,    
the amplitude depends on the four transverse degrees of freedom and rapidity, $Y=\ln(1/x)$,    
playing the role of the  evolution parameter    
\be    
\label{eq:NDef}    
N(\xb,\yb,Y)\,\equiv\, N_{\xb\yb}(Y)\,.   
\ee    
{From now on,    
for shortness of the notation,  we assume the $Y-$dependence implicit.}

In the leading logarithmic approximation, the dipole    
scattering amplitude  obeys a nonlinear evolution equation     
derived by Balitsky and Kovchegov \cite{BAL1,KOV1}    
\be    
\label{eq:BK}    
\frac{\partial N_{\xb\yb}}{\partial Y}\,=\,    
\overline{\alpha}_s\!    
\int\frac{d^2\zb}{2\pi}    
\frac{(\xb-\yb)^2}{(\xb-\zb)^2(\yb-\zb)^2}\,    
\left\{    
N_{\xb\zb}+N_{\yb\zb}-N_{\xb\yb}    
- N_{\xb\zb}\,N_{\yb\zb}   
\right\}\, ,    
\ee    
where $\overline{\alpha}_s=\alpha_s N_c/\pi$. In addition,    
one has to specify  an initial condition at $Y=Y_0$: $N_{\xb\yb}=N^0(\rb,\bb)$.    
The amplitude $N(\xb, \yb)$ in (\ref{eq:BK}) is given by the following correlator    
\be   
\label{eq:Corr}   
N(\xb,\yb)\, = \,\frac{1}{N_c}\mbox{\rm Tr} \left<1-U^{\dagger}(\xb) U(\yb)\right> \, ,   
\ee   
where the trace is done in the colour space,  and   
the eikonal factor $U$ is defined as the path ordered exponential with the $SU(N)$   
gauge fields (in the gauge $A_a^{-}=0$)   
\be   
\label{eq:eikonal}   
U(\xb)\, = \,  P\,  \exp\left\{i\int dx^{-}\,  T^a A_a^+(x^{-},\xb)\right\}\,.   
\ee   
The averaging $\left<...\right>$ in (\ref{eq:Corr}) is performed over an ensemble of classical gauge fields.   
In general, an infinite hierarchy of equations is found for correlators of the $U$ factors   
\cite{BAL1}. In the large $N_c$ limit, however, the closed form (\ref{eq:BK}) can be found   
for the two point amplitude $N_{\xb\yb}$ \cite{KOV1}.   
The linear part of (\ref{eq:BK}) corresponds to the  dipole version \cite{Mueller:94} of the   
BFKL equation \cite{Lipatov:86} at nonzero impact parameter and its solution has been studied in  
the Monte Carlo simulation  of onium-onium scattering \cite{Salam:96,Salam:97}. 
The additional quadratic term  emerges due to the summation  of the multiple interactions  
of the dipoles in the quark-antiquark wave function with the nucleus.    
    
\begin{figure}[htb]    
\centering\includegraphics[width=0.3\textwidth]{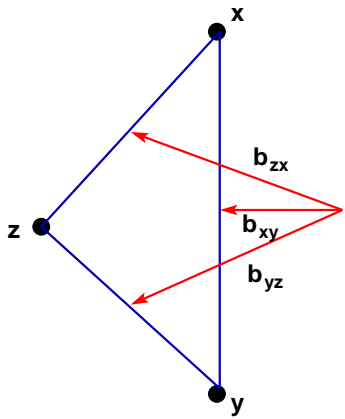}    
\caption{The triangle geometry in Eq.~(\ref{eq:BK}). The points   
$\xb,\yb,\zb$ are the dipole ends in the transverse space, and the    
vectors $\bb_{\xb\zb},\bb_{\xb\yb},\bb_{\zb\yb}$ are the impact parameters of the three dipoles.}   
\label{fig:triangle}   
\end{figure}

The r.h.s of Eq.~(\ref{eq:BK}) has a nice    
geometrical interpretation: the parent dipole $(\xb,\yb)$ splits into two    
new dipoles $(\xb,\zb)$ and  $(\yb,\zb)$, and the summation is taken over all    
new dipoles. On the other hand, the nonlinear term describes  the recombination of the two dipoles     
$(\xb,\zb)$ and  $(\yb,\zb)$  into one $(\xb,\yb)$.     
The three dipoles form a triangle, shown in Fig.~\ref{fig:triangle}.    
The first part of the integral kernel in (\ref{eq:BK}) only depends  on the triangle sides:    
$|\xb-\yb|$, $|\xb-\zb|$ and $|\yb-\zb|$. A nontrivial dependence on    
the position of the triangle in the plane, i.e. the dependence on the impact parameter vectors   
$\bb_{\xb\yb},\bb_{\zb\xb},\bb_{\yb\zb}$, is    
introduced through the arguments of the amplitudes $N$.   
In particular, it is interesting to study how    
the  $b$-dependence introduced by an initial condition $N^0(\rb,\bb)$ propagates with increasing $Y$.    
Let us also note that the singularities at $\zb=\xb,\yb$ in Eq.~(\ref{eq:BK}) are integrable provided    
\be    
\label{eq:2.3}    
\lim_{\xb\to \yb} N_{\xb\yb}\,\sim\,|\xb-\yb|^\epsilon\,,~~~~~~~\epsilon>0\,.    
\ee    
Thus  the dipole which shrinks to a point does not scatter.    
    
\subsection{Symmetries of the BK equation}

The BK equation (\ref{eq:BK}) has a rich symmetry structure. Introducing the complex number notation    
for the transverse vectors, e.g. $x=x_1+i\, x_2$ and $\overline x=x_1-i\, x_2$   
for $\xb=(x_1,x_2)$, it can be easily shown that   
the measure in Eq.~(\ref{eq:BK}),   
\be   
\label{eq:Measure}    
\frac{\asb}{2\pi}\,\,\frac{(\xb-\yb)^2}{(\xb-\zb)^2(\yb-\zb)^2}\,d^2\zb\,,    
\ee   
is invariant under the M\"obius transformation\footnote{Provided $\asb$ is kept constant.}   
\be   
\label{eq:Mobius}   
x\,\rightarrow\, \frac{a\, x+b}{c\,x+d}\,,~~~~~~~~\overline x\,\rightarrow\,    
\frac{\overline a \,\overline x+\overline b}{\overline c\, \overline x+\overline d}\,,   
\ee   
where the parameters  $a,b,c,d \in {\mathbb C}$  and  
$ad-bc \ne 0$.    
Identical transformations   
are also applied to $y\,(\overline y)$ and $z\,(\overline z)$.   
Thus the BK equation is {\it covariant} with    
respect to the M\"obius transformation.   
In particular, the following  elementary  
 transformations from (\ref{eq:Mobius}) are relevant for our discussion   
\begin{itemize}    
\item[--]  global two-dimensional translations  by  vectors $\bb$:~~~~~$\xb\, \to\, \xb+\bb$\,,   
\item[--]  global two-dimensional rotations by angles $\phi$:~~~~~$\xb\, \to\, O(\phi)\, \xb$\,,   
\item[--]  scale transformations with a real, positive parameter $\lambda$:~~~~~$\xb\, \to\, \lambda\,\xb$\,   
\item[--]  inversion (in complex notation):~~~~~$x\,\to\,1/x$\,.    
\end{itemize}   
   
Notice, that if an initial condition $N^0$ is invariant under any of the discussed transformations,   
the solution of the BK equation $N_{\xb\yb}(Y)$ preserves the corresponding symmetry.    
In particular, if an initial condition is invariant    
under translations and rotations, $N(\xb,\yb)=N^0(|\xb-\yb|)$,    
the solution at any rapidity $Y$ has the same property, i.e. it only depends on   
the dipole size $r=|\xb-\yb|$ but not    
on the impact parameter $\bb=(\xb+\yb)/2$.   
The problem of finding solution simplifies    
enormously in this case since only one degree of freedom is relevant,   
namely the dipole size $r$.   
   
Physically, this approximation (called {\it local approximation})   
corresponds to an infinitely large and uniform nucleus.    
Previous analytical \cite{KOV2,LEVTUCH} and numerical \cite{BRAUN,LEVLUB,GBMS}    
studies of  the BK equation were based upon this assumption.     
In this approximation, the solution shows saturation, $N(r)\to 1$, with    
the characteristic scale $Q_s(Y)$.    
For dipoles smaller than the inverse of the saturation scale, $r<1/Q_s(Y)$, the solution is governed    
mainly by the linear term of Eq.~(\ref{eq:BK}) and shows the exponential rise in rapidity, $N\sim \exp(\omp Y)$  
where $\omp=4\ln 2\, \asb$ is the intercept of the BFKL kernel.    
On the other hand, in the region where dipoles are large,  $r>1/Q_s(Y)$, the nonlinear term   
slows down the rise      
and eventually the amplitude saturates to 1.   
The saturation scale $Q_s(Y)$ depends on the rapidity in the following way \cite{KOV2,LEVTUCH,BRAUN,GBMS,BL}   
\be    
\label{eq:SatScaleOld}    
Q_s(Y)\, =\, Q_0 \exp({\lambda\, \asb} Y)  \, ,   
\ee    
where the coefficient\footnote{To be precise in Ref.~\cite{KOV2} the coefficient was found to be the same as  
the Pomeron intercept $\lambda=\omp/\asb=4\ln2$, but this value was
not confirmed by subsequent analytical and numerical studies.} $\lambda \simeq 2$.    
The solution in the local approximation also exhibits a property of the {\it  geometric scaling} \cite{SGK},    
namely for $r>1/Q_s(Y)$   
\be    
\label{eq:GS}   
 N(r,Y) \,  \equiv \, N(r Q_s(Y)) \, ,    
\ee    
which means that the amplitude $N$ in the saturated region only depends on one combined variable    
$r Q_s(Y)$ instead of $r$ and $Y$ separately\footnote{This is also a feature of the saturation model \cite{GBW}.}.    
Thus the  diffusion into infrared, typical for the linear BFKL equation,    
is damped by the emergence of the saturation scale $Q_s(Y)$ \cite{GBMS}.    
    
\begin{figure}[htb]    
\centering\includegraphics[width=0.3\textwidth]{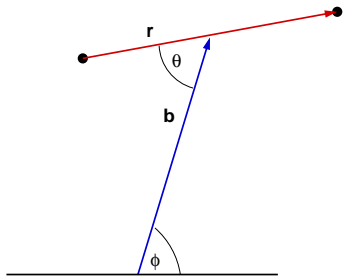}    
\caption{Dipole degrees of freedom: the dipole vector $\rb$ and the impact parameter   
$\bb$.}    
\label{fig:dippos1}    
\end{figure}    
    
Since in our analysis we want to study the impact parameter dependence, we    
obviously have to abandon the assumption about the translational invariance.    
However, in order to simplify the problem, we adopt more physical assumption that our nucleus    
is cylindrically  symmetric, i.e. $N(\xb,\yb)$ is    
 invariant under the global rotation.    
It means that in the parameterisation    
of the dipole position,   
see Fig.~\ref{fig:dippos1},     
\be    
\label{eq:ParamDip}    
(\xb,\yb)\,=\,(r,b,\theta,\phi)\,,    
\ee    
we drop the dependence on the azimuthal angle $\phi$.    
Note that we keep the dependence on     
$\theta$ which  is the angle between the vectors $\rb$ and $\bb$    
\be    
\cos\theta\,=\,\frac{\rb\cdot \bb}{r\,b}\,.    
\label{eq:costheta}    
\ee    
The assumption about cylindrical symmetry     
reduces the number of parameters in the amplitude $N$ to four: three degrees of freedom for a dipole    
and the evolution variable $Y$.

\section{The numerical method of finding solution}    
   
The numerical method for the solution of the BK equation    
is very similar to the one used for the solutions of the linear equations, see \cite{BoMaSaSc:97}.   
We discretise the amplitude $N(r,b,\cos\theta)$ in all 3 variables   
$(\ln r, \ln b , \cos\theta)$. A   
simple linear interpolation has been chosen which is the fastest in this case.    
To find the solution as a function of rapidity we take in the first step $N^0$, the initial condition at $Y=Y_0$,    
and evaluate  Eq.~(\ref{eq:BK}) with the step $\Delta Y$. This gives the first approximation   
for the solution at $Y=\Delta Y$
\be   
\label{eq:1st}   
N_{\xb\yb}^{(1)}\,=\,N_{\xb\yb}^0+\Delta Y \int_{\zb}\, \{   
N_{\xb\zb}^0+N_{\yb\zb}^0-N_{\xb\yb}^0-N_{\xb\zb}^0N_{\yb\zb}^0   \; ,
\}\,   
\ee   
where we symbolically denoted the integration over ${\zb}$ with the measure (\ref{eq:Measure}).   
If the relative difference $|(N_{\xb\yb}^{(1)}-N_{\xb\yb}^0)/N_{\xb\yb}^0|<\epsilon$ (some    
accuracy) at each point of the grid,  we finish our procedure, otherwise   
solution (\ref{eq:1st}) is used    
to find the second approximation from:   
\beeq   
\label{eq:2st}   
\nonumber   
N_{\xb\yb}^{(2)}\,=\,N_{\xb\yb}^0\!\!\!&+&\!\!\!\frac{\Delta Y}{2} \int_{\zb}\, \left[   
N_{\xb\zb}^0+N_{\yb\zb}^0-N_{\xb\yb}^0-N_{\xb\zb}^0 N_{\yb\zb}^0   
\right]   
\\ \nonumber   
\\ \nonumber   
\!\!\!&+&\!\!\!\frac{\Delta Y}{2}  \int_{\zb}\, \left[   
N_{\xb\zb}^{(1)}+N_{\yb\zb}^{(1)}-N_{\xb\yb}^{(1)}   
-N_{\xb\zb}^{(1)} N_{\yb\zb}^{(1)}   
\right]   
\\ \nonumber   
\\   
\!\!\!&+&\!\!\!\frac{\Delta Y}{6} \int_{\zb}\, \left[   
N_{\xb\zb}^{(1)}-N_{\xb\zb}^0   
\right]   
\left[   
N_{\yb\zb}^{(1)}-N_{\yb\zb}^0   
\right]\,.   
\eeeq   
The r.h.s. of Eq.~(\ref{eq:2st}) was found after integrating over $Y$ from $0$ to $\Delta Y$,   
assuming a linear interpolation in $Y$ between $N^0$ and $N^{(1)}(\Delta Y)$. This is justified for small   
enough value of $\Delta Y$. We iterate in this way, using Eq.~(\ref{eq:2st}),    
until the desired accuracy is achieved.

Usually only couple of iterations are needed to   
find the right answer, in fact we fix the maximal number of iterations to $5$.   
In order to get satisfactory results one has to work with    
a grid which is at least $(100_r \times 100_b \times 20_c)$.    
Each step in rapidity produces  about $1.5~\mbox{\rm MB}$    
of output data and takes about $300~\mbox{\rm min.}$  of CPU time when run on     
PC machine with $2.5~\mbox{\rm  GHz}$ processor and $2.0~\mbox{\rm GB}$ RAM memory.

One of the important issues while studying the BK equation is the choice    
of the initial condition  at rapidity $Y=0$. As mentioned above, because of    
the cylindrical symmetry our initial amplitude $N^0$ should only depend on the three variables   
$(r,b,\theta)$. Since we have nearly no information    
on the angle  $\theta$ between $\rb$ and $\bb$, we do not assume any  dependence on it     
in the initial conditions.    
We shall see  that the nontrivial $\theta$ dependence is nevertheless generated through the evolution.    
As far as the $r$ and $b$ dependence is concerned we have chosen the initial distribution in the    
Glauber--Mueller form \cite{GLAUBER,MUELLERSAT}    
\be    
N^0(r,b) = 1-\exp\{-r^2 S(b)\}\; ,   
\label{eq:GM}   
\ee   
with $S(b)$ being a steeply falling profile in $b$, e.g. $S(b) \sim \exp(-b^2)$.   
The Glauber--Mueller formula resums the multiple scatterings of a single dipole on a nuclear target   
and it has been advocated to be a natural choice for the starting distribution of the BK equation \cite{KOV1}.     
Let us note that the above distribution has a property that for any fixed value of impact parameter    
it saturates to $1$ for sufficiently large values of the dipole size $r$.   
More detailed analysis with other forms of the initial conditions will    
be presented elsewhere \cite{GBS2}.   
   
\section{The dependence of the solution on a dipole size $r$}   
   
\begin{figure}[t]    
\centering\includegraphics[width=0.47\textwidth]{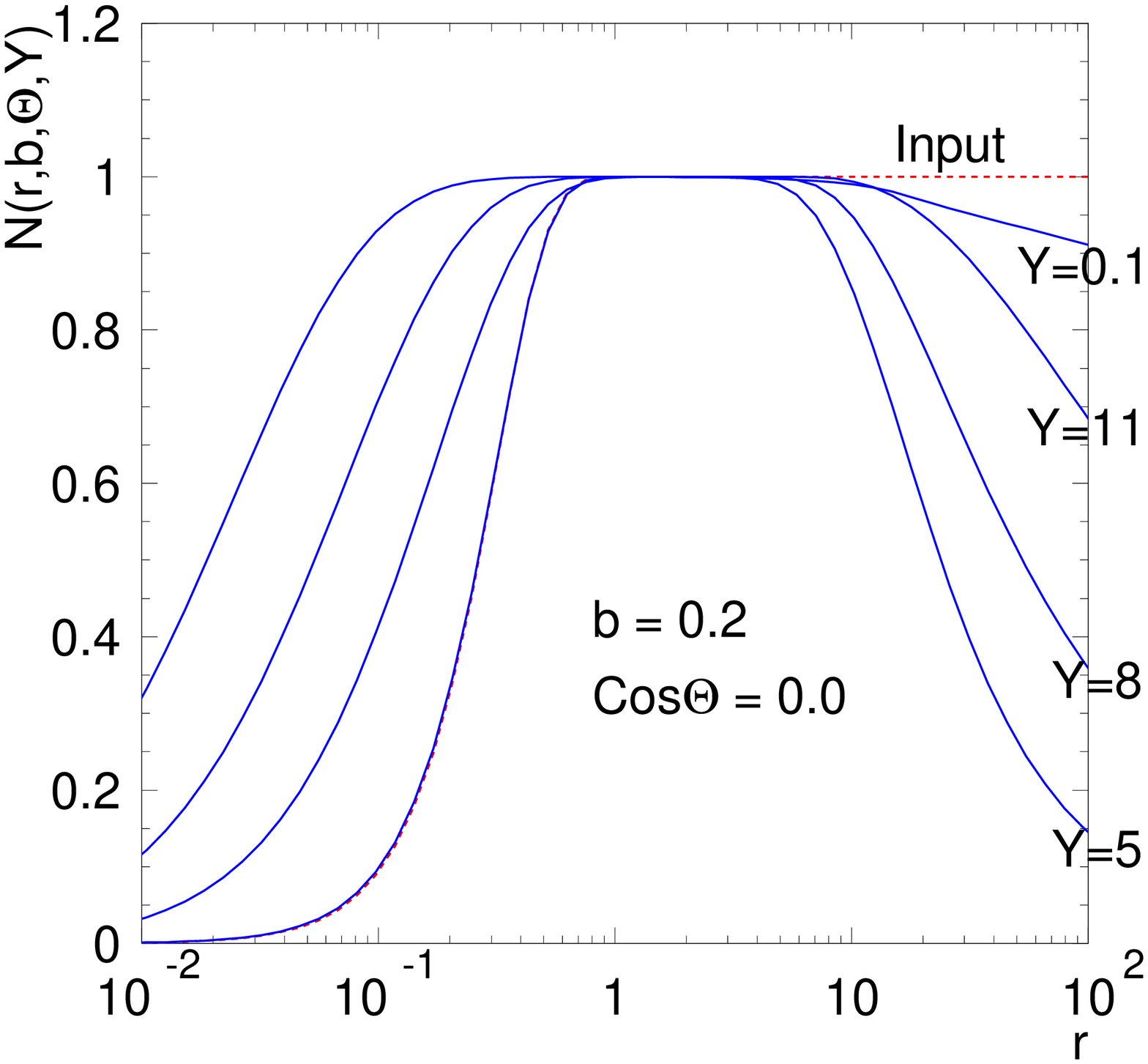}  \hfill   
\includegraphics[width=0.47\textwidth]{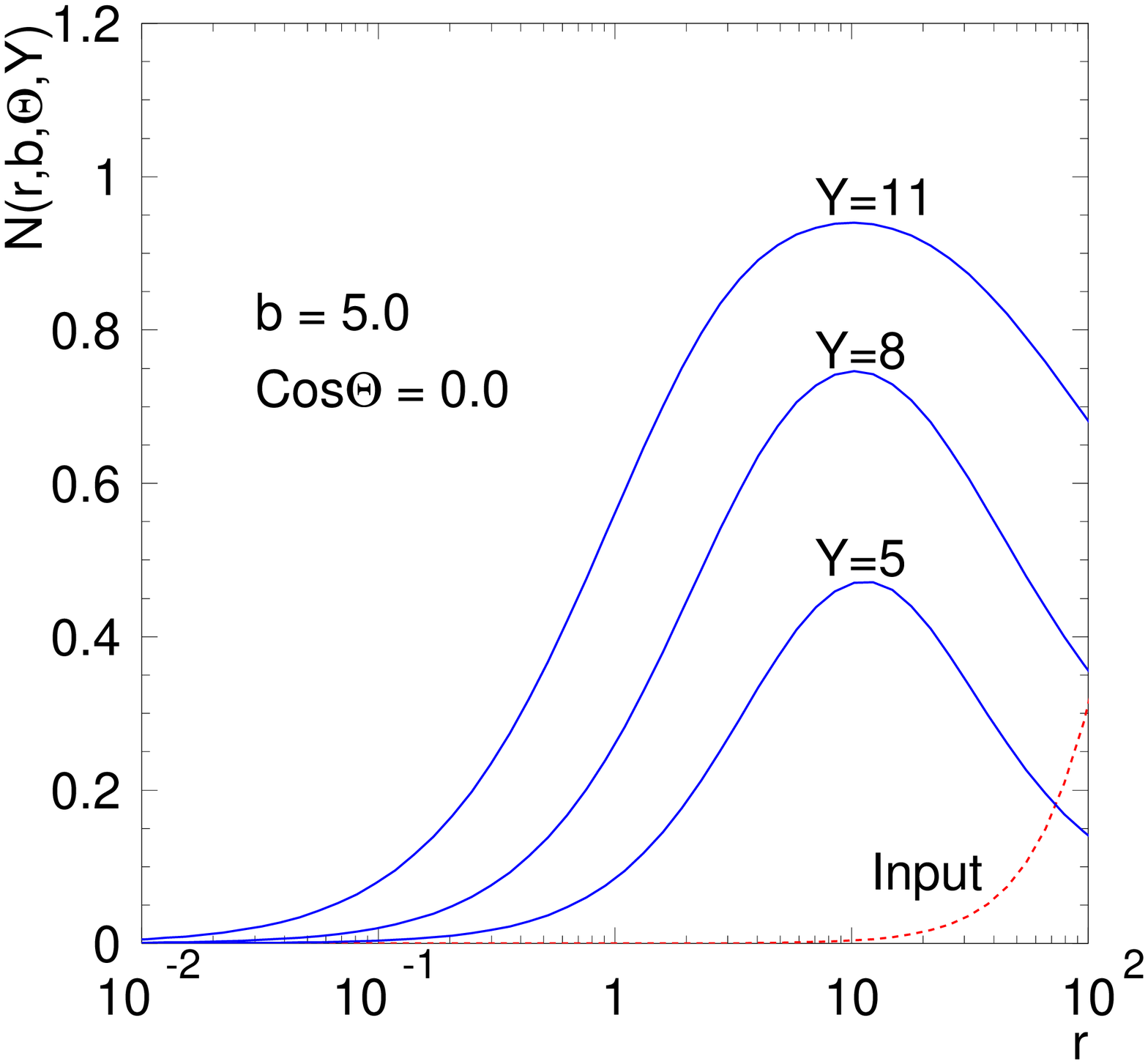}   
\caption{The amplitude $N(r,b,\theta;Y)$ as a function of dipole size $r$    
for indicated values of rapidity $Y$ and two values of the fixed impact   
parameter: $b=0.2$ (left) and $b=5$ (right). The orientation of the dipole   
$\cos\theta=0$. The red-dashed line is the input distribution (\ref{eq:GM}) with profile (\ref{eq:Sb}).   
}   
\label{fig:rdepb}    
\end{figure}    
   
We have performed the numerical evolution of the BK equation starting from   
formula (\ref{eq:GM}) as the initial condition with the following impact parameter profile   
\be   
S(b)=10\, \exp\{-b^2/2\} \,.   
\label{eq:Sb}   
\ee   
Throughout this work we keep the coupling constant in (\ref{eq:BK}) fixed, $\asb=0.2$.   
The running of the coupling, although physically more justified, would be   
an additional complication to the solution which we would like to avoid at this stage\footnote{   
In the local approximation,   
running coupling changes rapidity dependence of the saturation scale:   
$Q_s^{\rm run}(Y) \sim \exp(c\sqrt{Y})$,  see \cite{GBMS,IaItMcL:02,MuTr:02}
and also \cite{GLR}.}. 
Nevertheless, one should stress that a running coupling is an important NLL effect \cite{BFKLNEXT},  
which should be taken into account in future investigations.  
We are going to study now how such  initial condition (\ref{eq:GM})    
is modified by the evolution in rapidity. We concentrate first on the   
dependence of the solution on a dipole size $r$.

The results of our numerical analysis are shown in Fig.~\ref{fig:rdepb} for two values   
of the impact fixed impact parameter: $b=0.2$ and $b=5$.  We have also fixed the orientation of the dipole such that   
$\rb \perp \bb$ that is $\cos\theta=0$, see Eq.~(\ref{eq:costheta}).   
As we see, for the values of $r$ which are small (here $r\ll 1$)    
the amplitude rises with increasing values of rapidity $Y$.   
In this region, the linear part of the BK equation dominates. With increasing $r$, for fixed    
rapidity, the  amplitude finally reaches  the saturation value $N_{\rm sat}=1$, were   
the nonlinearity of the  BK equation is crucial.   
This is mostly visible for small value of impact parameter, Fig.~\ref{fig:rdepb} (left).  
For a larger value of   
impact parameter, $b=5$, in the plot to the right, the saturation has been only reached for  
the highest value of rapidity shown $Y=11$.  
What is interesting is the fact that the amplitude has a maximum for the dipole size  
which is twice its impact parameter $r=2 b$, that is $r=10$ in this case. 
   
At large $r$, the behaviour of the amplitude is quite different from the case with    
the translational invariance (no $b-$dependence).    
The amplitude  decreases with increasing $r$ even if the  initial condition   
is saturated ($N^0=1$) in this region. The reason for this behaviour can be easily understood   
analysing the  solution after the first iteration (\ref{eq:1st})  
\be   
\label{eq:s1}   
N^{(1)}_{\xb\yb}\,=\,N^{0}_{\xb\yb}\,+\,\Delta Y\,   
\frac{\overline{\alpha}_s}{2\pi}\!   
\int{d^2\zb}\,   
\frac{(\xb-\yb)^2}{(\xb-\zb)^2(\yb-\zb)^2}\,    
\left\{  
N^0_{\zb\yb}+N^0_{\xb\zb}-N^0_{\xb\yb} -N^{0}_{\xb\zb}\, N^{0}_{\yb\zb}   
\right\}   
\, ,   
\ee   
with the Glauber-Mueller initial condition   
\be   
\label{eq:s0}   
N^{0}(r,b)\,=\,1-\exp\{-c\,r^2\exp(-b^2 d)\}\,.   
\ee   
Saturation means that for a large enough $r=|\xb-\yb|$ (and fixed $b$)   
$N_{\xb\yb}\to 1$, i.e.   
a large dipole is totally absorbed. This is indeed the case for the initial condition (\ref{eq:s0}).   
After the first evolution step, however, $N^{(1)}_{\xb\yb}$ becomes less than $1$  for large $r$ 
due to the large and negative integrand in the curly brackets in (\ref{eq:s1}), which    
leads to the effect shown in Fig.~\ref{fig:rdepb} at large $r$. 
   
Indeed, if $N^0_{\xb\yb} \simeq 1$ in Eq.~(\ref{eq:s1}), the integrand is always negative or equal zero 
\be 
N^0_{\xb\zb}+N^0_{\yb\zb}-1-N^0_{\yb\zb}N^0_{\xb\zb} \, = \, -(1-N^0_{\zb\yb})(1-N^0_{\xb\zb}) \, \le \, 0\,.  
\label{eq:integrand} 
\ee 
In Fig.~\ref{fig:triangle} we show an example of 
typical configurations of the $(\xb,\zb)$ and $(\yb,\zb)$ dipoles which give dominant contribution 
to the integral in Eq.~(\ref{eq:s1}). For such a  contribution 
$r_{xz}\approx r_{yz}\approx 2 b_{xz}\approx 2b_{yz}\approx r$ (for $b_{xy}\approx 0$), and  
for the initial condition (\ref{eq:s0}) the amplitudes  
$N^{0}_{\xb\zb}\approx N^{0}_{\yb\zb}\approx 0$ for the sufficiently
large value of $r$. Thus  
the integral picks up significant negative contribution from this configuration.   
This should be contrasted to the translationally invariant case with no $b-$dependence   
in the initial condition, e.g. $N^{0}(r)=1-\exp\{-r^2\}$. For such  initial condition the  
 configuration from Fig.~\ref{fig:triangle} gives $N^{0}_{\xb\zb}\approx N^{0}_{\yb\zb} \simeq 1$, and 
the expression in  Eq.~(\ref{eq:integrand}) is negligible. One can prove that any other configuration,  
for example when $|\xb-\zb| \ll |\xb-\yb|$ or   
$|\xb-\zb| \gg |\xb-\yb|$, leads to the same result. Thus, in the case of   
infinitely large and uniform nucleus, the amplitude $N$ stays always saturated for large dipole sizes.   
   
In the next evolution step, $N_{\xb\yb}^{(1)}$ plays the role of the initial condition.   
Since $N_{\xb\yb}^{(1)}$ is no longer equal to $1$ for large dipole sizes, the analysis becomes more complicated.  
At some rapidity, the discussed integral becomes positive  
leading to the effect observed in Fig.~\ref{fig:rdepb} where $N$ starts to grow again, reaching unity at large $r$  
for  large values of rapidity.    
   
In more physical terms the fall-off of the amplitude at large $r$ can be explained by realizing that   
the end points $\xb$ and $\yb$  of    
a sufficiently large dipole are in the region where there is no   
gauge field. In this case $U(\xb)=U(\yb)=1$, and the correlator  (\ref{eq:Corr}) vanishes.    
It simply means that the dipole is so large that it misses the localised target given by $S(b)$.    
On the other hand, the  non-vanishing value of the amplitude for very large values of $r$     
in the previous studies in the local approximation \cite{KOV2,LEVTUCH,BRAUN,LEVLUB,GBMS}    
was a consequence of the fact that the field was uniform and present everywhere,    
thus it did not matter how large the dipole was, it always scattered. The appearance of the gauge field,   
signalled by the increase of N with rising rapidity, corresponds to expansion of a black (or grey) region   
of a nucleus, see section \ref{sec:black}.

\subsection{Saturation scale and geometric scaling}   
\label{subsec:satscale}
The saturation scale $Q_s$ is defined from the condition   
\be   
\label{eq:satscb}   
\left<\,N(r=1/Q_s,b,\theta,Y)\,\right>_{\theta}\,=\,\kappa  \, , 
\ee   
where the constant $\kappa$ is of the order of unity ($\kappa=1/2$ in our numerical analysis)  
and we have taken an average over the orientation of the dipole with respect to the vector of impact parameter.   
Notice that after solving the above equation, the saturation scale depends on impact parameter   
in addition to rapidity\footnote{Of course, $Q_s$ depends also on $\kappa$ but this dependence is   
irrelevant for our discussion.}. For the form of $N$, shown in Fig.~\ref{fig:rdepb},  there are   
two solutions of  Eq.~(\ref{eq:satscb})   
for sufficiently large rapidities, corresponding to small and large dipole sizes. Thus saturation   
(defined here as $N>\kappa$) occurs over a finite region of dipole sizes   
\be   
\label{eq:TwoScales}   
1/Q_s(b,Y) < r < R_{H}(b,Y) \,.   
\ee   
   
\begin{figure}[h]    
\centering\includegraphics[width=0.7\textwidth]{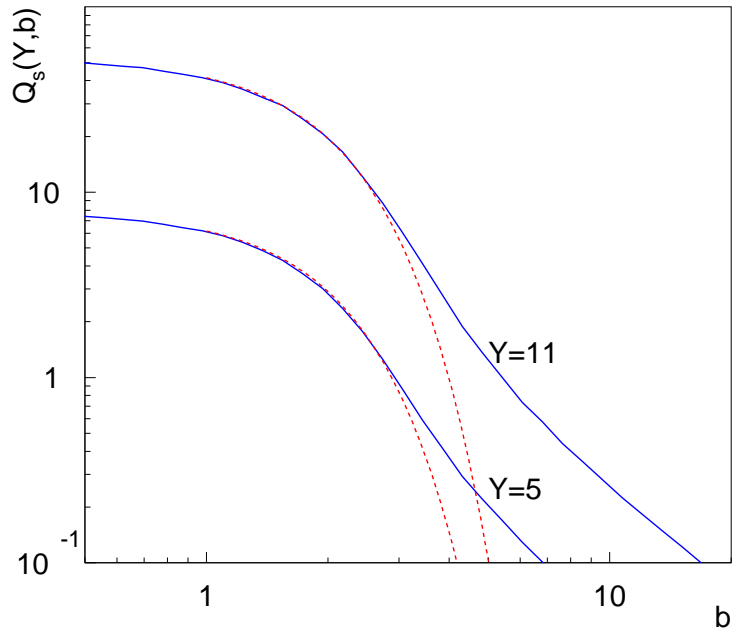}    
\vspace{-1cm}   
\caption{The dependence of the saturation scale $Q_s(b,Y)$ on impact parameter   
for two fixed values of rapidity. For the comparison  the  saturation scale at initial rapidity (\ref{eq:SatIni}) 
is shown by the dashed line, normalised to the value of $Q_s(b,Y)$ at $b=0.1$.}   
\label{fig:scb}   
\end{figure}

The scale $Q_s(b,Y)$ is the $b$-dependent saturation momentum  introduced by Mueller \cite{MUELLERSAT}.   
The $b-$independent saturation scale  found in    
the previous analyses \cite{LEVTUCH,BRAUN,LEVLUB,GBMS} can be regarded as an average over  
all area of interaction $Q_s(Y) =\left<Q_s(Y,b)\right>_b$. The second scale $R_H(b,Y)$    
appears because we have introduced exponential impact parameter profile,    
as has been explained in detail in the previous section, and reflects the boundary of   
the nucleus.   
In Fig.~\ref{fig:scb} we show the form of the impact parameter profile of   
the saturation scale $Q_s(b,Y)$ for two different rapidities $Y=5$ and $Y=11$.  
The saturation scale at $Y=0$, shown by the dashed line for comparison,    
is computed from    
the initial condition (\ref{eq:GM}) with (\ref{eq:Sb})
\be    
\label{eq:SatIni}    
Q^2_s(b,0)\,=\,Q_0^2\, \exp(-b^2/2)\,,   
\ee    
where we have fixed  the normalisation $Q_0$ to match the $Q_s(b,Y)$ at small values of $b=0.1$. 
Thus the dashed lines in Fig.~4, show the profile of $Q_s(b,Y)$ which we would get  
if the exponential impact parameter dependence set in the initial conditions would be preserved through the evolution. 
We clearly see that for higher impact parameters, the exponential   
fall-off set by initial condition  is replaced by the power-like tail
$Q_s(b,Y=11) \sim 1/b^{\gamma}$  
with $\gamma \simeq 1.6-2.0$ for $b>7$.   
   
As it is evident from Fig.~\ref{fig:rdepb}, both the saturation scale $Q_s(b,Y)$    
and $R_H(b,Y)$ rise with increasing rapidity, thus leading to broadening of    
the region in $r$ for which saturation occurs.   
The energy dependence of $Q_s(b,Y)$ has been estimated to be proportional to  $\exp(\lambda_s \asb Y)$    
with $\lambda_s \simeq 1.7 - 2.0$ at the highest  values of rapidity $Y=9-11$. 
We also checked that this value of $\lambda_s$ is not very sensitive to impact parameter. 
This would mean that (at high rapidities) the saturation scale has a factorised form  
$Q_s(Y,b) = \exp(\lambda_s \asb Y)\, g(b)$. 
We expect exponential dependence on rapidity   
of $R_H(b,Y)\sim \exp(\lambda_H \asb Y)$ too,   
and we have found  $\lambda_H \simeq 2.9$.

\begin{figure}[h]    
\centering\includegraphics[width=0.7\textwidth]{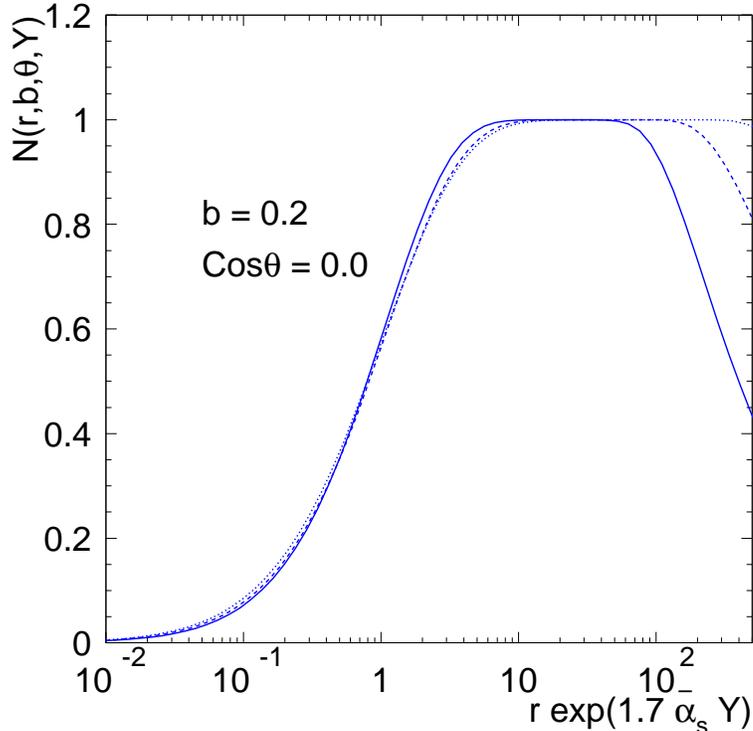}    
\vspace{-1cm}   
\caption{Geometric scaling for $b=0.2$   
obtained after rescaling $r$ by the rapidity dependence of the saturation scale   
$Q_s(b,Y)$. Dotted, dashed and solid lines are for $Y=11,9,7$ respectively.  
The orientation of the dipole has been fixed such that $\cos \theta=0$.}   
\label{fig:geoscal}   
\end{figure}

Since there are two scales in the problem, $Q_s$ and $R_H$, geometric scaling of the form (\ref{eq:GS})    
is not present anymore for arbitrarily large dipoles.   
However, when $r \ll R_{H}(b,Y)$ there is still an approximate    
geometric scaling in the combined variable $r \,Q_s(b,Y)$ for fixed $b$.  
This is shown in Fig.~\ref{fig:geoscal}, obtained from   
Fig.~\ref{fig:rdepb} after rescaling $r$ by $\exp(1.7\, \asb Y)$, i.e. by the rapidity dependence   
of the saturation scale $Q_s(b,Y)$, extracted in the region $7 \le Y \le 11$.   
As discussed, there is no geometric scaling for large dipole sizes. The problem of scaling in the    
$b-$dependent case has already been addressed in \cite{FIIM}    
and in recent phenomenological studies \cite{MuWal:03}, and still needs deeper analysis.

Let us finally note that for our solution there is a region in $b$ in   
 which the saturation scale   
does not exist at all. It is evident from the right hand side plot in Fig.~\ref{fig:rdepb}, where   
up to the rapidities $Y \sim 6$ the amplitude $N<\kappa$ for all values of the dipole sizes $r$,  
and Eq.~(\ref{eq:satscb}) does not have a real solution.

\section{Impact parameter dependence of the solution}  
  
We will discuss now the impact parameter dependence of the solution to  
the BK equation  
with the Glauber-Mueller input distribution (\ref{eq:GM},\ref{eq:Sb}).  
  
\begin{figure}[t]  
\centering\includegraphics[width=0.7\textwidth]{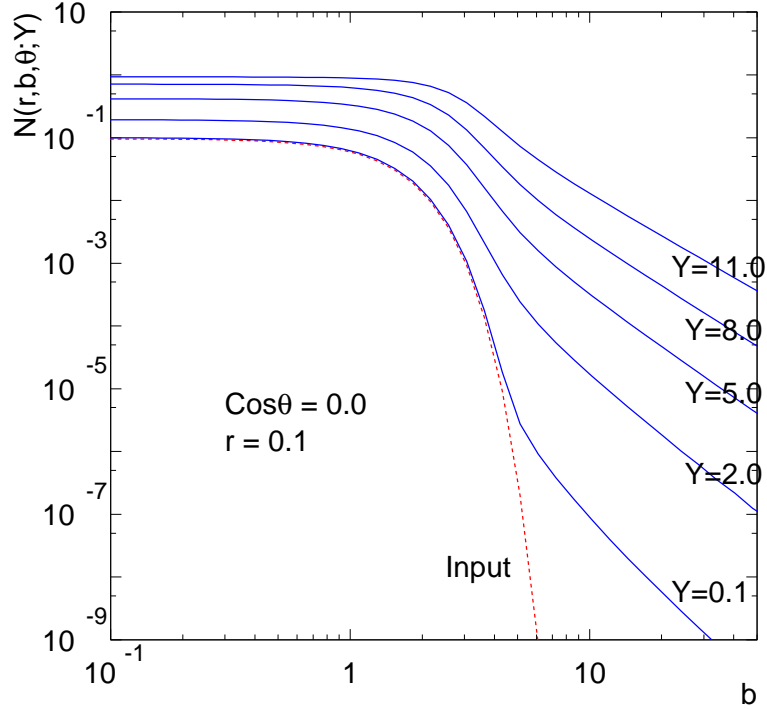}  
\vskip -1cm  
\caption{The amplitude $N(r,b,Y)$ as a function of impact parameter $b$  
for different values of rapidity $Y$.  
The dipole size and orientation are fixed, $r=0.1$ and $\cos\theta=0$.  
The dashed line is the input distribution (\ref{eq:GM}) with the profile  
(\ref{eq:Sb}).}  
\label{fig:bdep}  
\end{figure}

In Fig.~\ref{fig:bdep} we show the amplitude $N(r,b,\theta;Y)$ as a function  
of impact parameter $b$ for fixed and small  dipole size $r=0.1$ and  
various values of rapidity. We have also fixed the orientation of the  
dipole such that  
$\cos\theta=0$.  
The first striking feature of the solution is the fact that the steeply  
falling exponential dependence in $b$,  
given by the initial profile $S(b)$, is washed out by the evolution and  
instead  
clear power behaviour is generated for large values of $b$.  
Initially, at lowest rapidity $\Delta Y =0.1$ the amplitude $N(b)\sim  
1/b^{3.6}$, whereas later on  it  
has  a milder dependence: $N(b)\sim 1/b^\gamma$ with $\gamma$ between $2$  
and $3$ for $Y\ge 5$.  
The growth of the amplitude as a function of rapidity  
at large impact parameters is exponential,  
$N(Y)\sim\exp(\omega Y)$ with $\omega \simeq 2.7 \, \asb $  
for $b=10$. This strong dependence is clearly  
governed by the linear BFKL  part of the equation since the amplitude is  
small enough  
for the nonlinear term to be safely neglected and it is in a very good agreement 
with the hard Pomeron intercept  $\omp=4\ln 2\, \asb = 2.77\, \asb$.

On the other hand, at small values of impact parameters, $b<1$,  where the  
amplitude is large,  
we clearly observe that  the growth of amplitude is strongly damped due to  
the nonlinear term.  
This is the region of $b$ where the saturation sets in first.  
  
\subsection{The origin of the power-like tail}  
It has been argued by Kovner and Wiedeman in  \cite{KW}  that the power behaviour in impact  
parameter  
$N\sim 1/b^\gamma$ originates from the configurations with very large  
dipoles.  
Following \cite{KW} let us take  small dipole with size $r=|\xb-\yb|$  
which is  
located far away from the target, at large impact parameter $b$,  
in the area where the colour field is very weak. In that case one has  
$U(\xb) \simeq U(\yb) \simeq 1$  
and thus $N(\xb,\yb)\simeq 0$.  
  
The non-vanishing contribution to the r.h.s of equation (\ref{eq:BK}) comes  
from configurations of large dipoles  
with one end-point situated at $\xb$, and the other at $\zb$,  
close to the center of the target where the field is strong.  
The phase in eikonal (\ref{eq:eikonal}) oscillates strongly  and thus  
$\left<U(\zb)\right>\simeq 0$.  
Therefore $N(\xb,\zb)\simeq N(\yb,\zb) \simeq 1$, and this configuration  
gives  
\be  
\int d^2 \zb\, \frac{(\xb-\yb)^2}{(\xb-\zb)^2 (\yb-\zb)^2}\,  
(N_{\xb\zb}+N_{\yb\zb}-N_{\xb\yb} -N_{\xb\zb}N_{\yb\zb})\, \simeq\,  
\frac{r^2}{b^4}  
\int d^2 \zb \, \simeq \, \frac{r^2}{b^4}\, \pi R_0^2(Y)  
\label{eq:b4}  
\ee  
where $\pi R^2_0(Y)$ is the area of strong field in the target  over which  
we integrate.  
Strictly speaking the statement that $N(\xb,\zb)=1$ for the configuration  
considered above  
is valid only at very high rapidities. Due to the initial conditions  
(\ref{eq:GM})  
at intermediate rapidities there will be always such $b$, large enough,  
for which these configurations will have $N(\xb,\zb)<1$, however always  
$N(\xb,\zb) \gg N(\xb,\yb)$.  
  
\begin{figure}[t]  
\centering\includegraphics[width=0.7\textwidth]{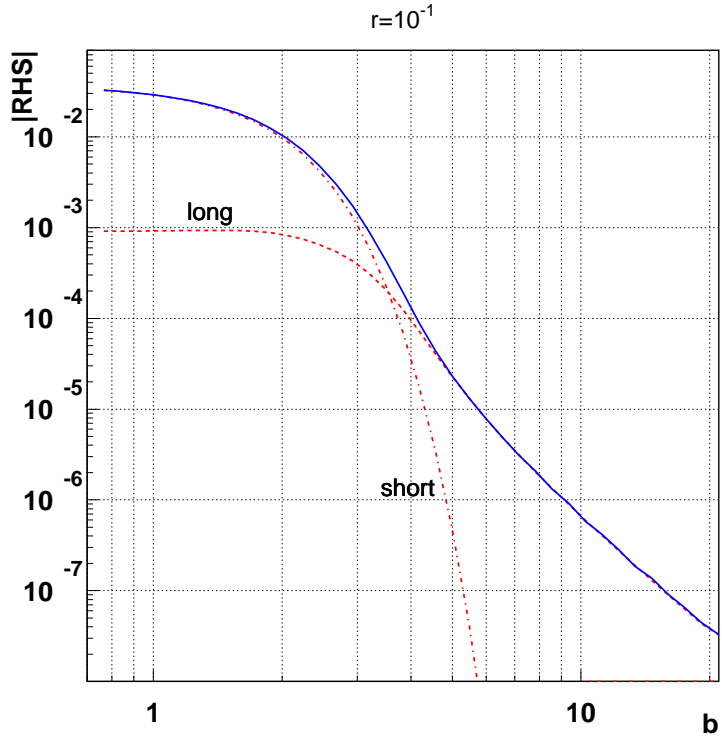}  
\vskip -0.5cm  
\caption{The r.h.s. of the BK equation (\ref{eq:BK})  
computed for the initial condition (\ref{eq:GM},\ref{eq:Sb})  
(solid line) and the contributions from the short and long  
dipoles (dashed lines), see (\ref{eq:ShortLong}),  
plotted as a function of $b$ for fixed dipole size $r=0.1$ and orientation  
$\cos\theta=0$.}  
\label{fig:sholo}  
\end{figure}  

In order to check this statement numerically  we  
 perform only one iteration of  
the BK equation  with very small step in rapidity $\Delta Y=0.1$,  
Eq.~(\ref{eq:1st}), and divide  
the integration region into two parts: $|\zb-\bb| < r_0$ and $|\zb-\bb| >  
r_0$  
with the cutoff $r_0 = 1$ for the choice of small dipole $r = 0.1$.  
To be precise we evaluate two  integrals:  
\be  
\bigg[\stackrel{short}{\overbrace{\int \Theta(r_0-|\zb-\bb|)}}  
+\stackrel{long}{\overbrace{\int \Theta(|\zb-\bb|-r_0)}} \bigg]  
\frac{d^2 \zb \, (\xb-\yb)^2}{(\xb-\zb)^2 (\yb-\zb)^2}  
(N^0_{\xb\zb}+N^0_{\yb\zb}-N^0_{\xb\yb} -  
N^0_{\xb\zb}N^0_{\yb\zb})  
\label{eq:ShortLong}  
\ee  
with $N^0$ being our initial condition.  
The results are presented in Fig.~\ref{fig:sholo}, where 
the short range contribution dominates at small values of impact  
parameters  
and the long range one appears to be responsible for the power behaviour  
in $b$ at large values.  
The point at which the long range contribution starts to dominate over the  
short range one  
is determined by the form of the initial condition.

\subsection{Angular dependence for large dipole sizes} 
\label{sec:angle} 
 
It is interesting to study also the angular dependence of the solution $N$. 
For sufficiently small dipole sizes $r \ll R_H$, the angular dependence is negligible.  
However, when $r \sim R_H$ the difference in the amplitude due to the dipole orientation is quite substantial,  
especially at large impact parameters $b \sim r$. 
 
\begin{figure}[t]    
\centering\includegraphics[width=0.47\textwidth]{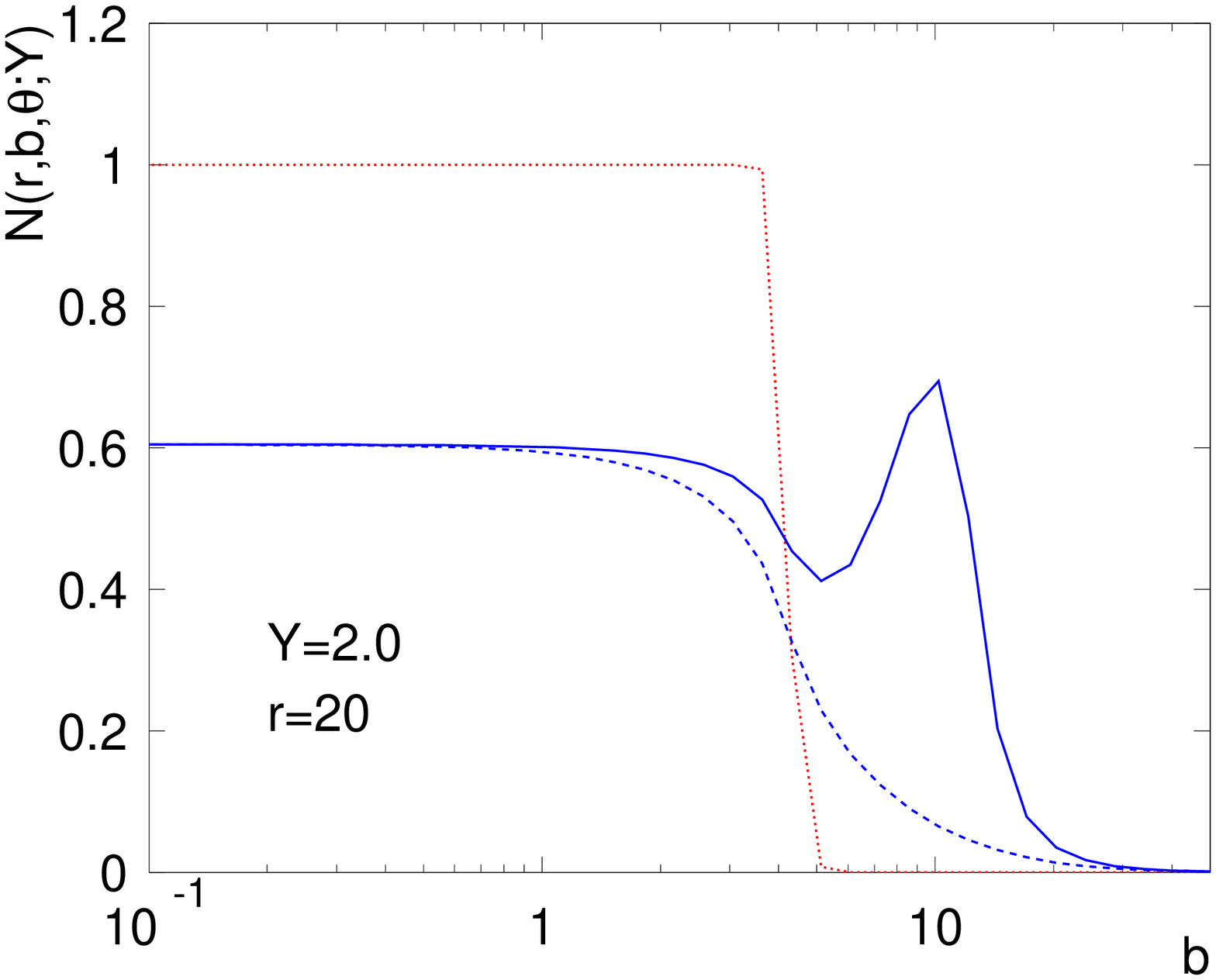}  \hfill   
\includegraphics[width=0.47\textwidth]{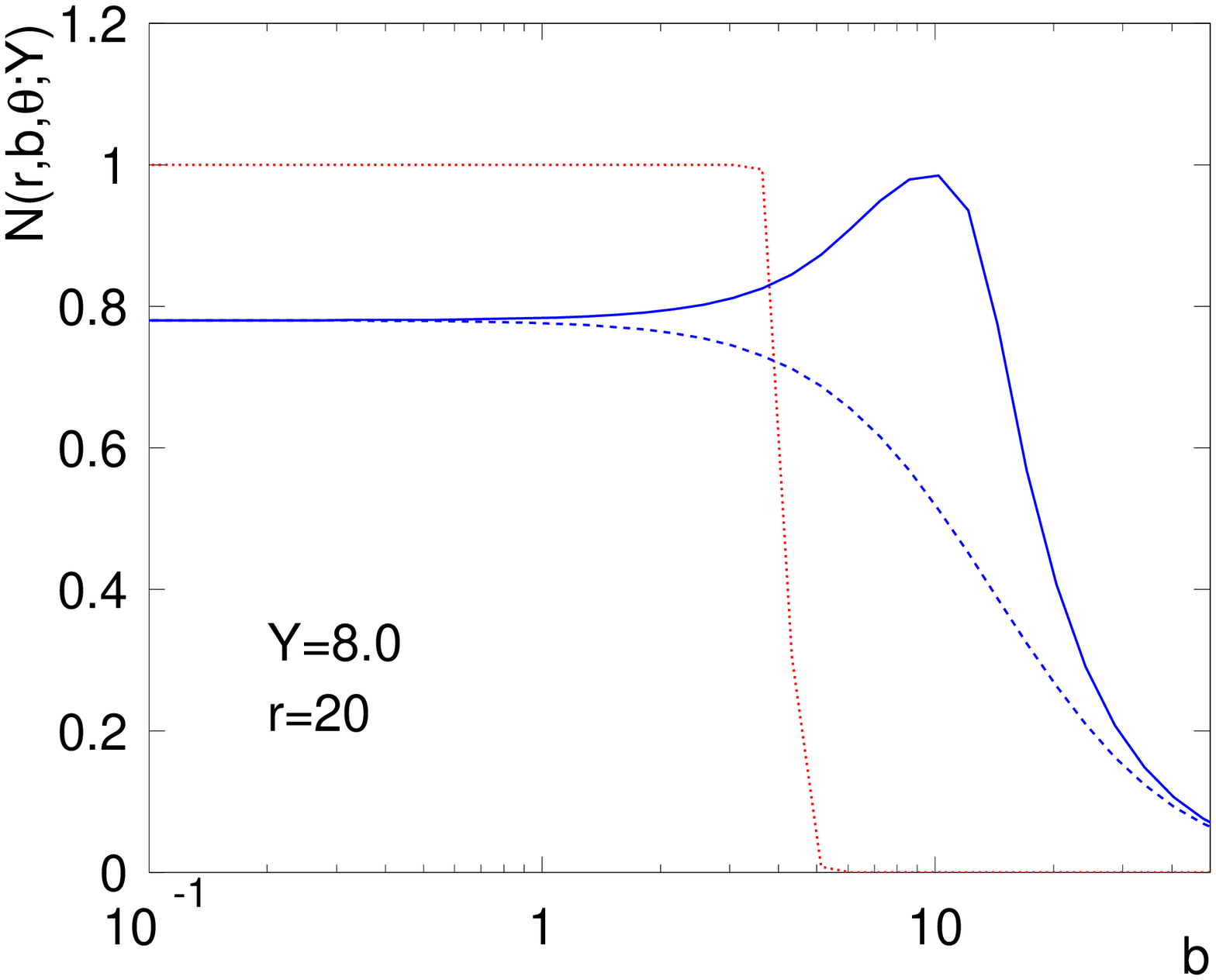}   
\caption{The amplitude $N(r,b,\theta;Y)$ as a function of impact parameter $b$    
for two rapidities:  $Y=2$ (left)  and $Y=8$ (right). 
The dipole size is fixed to $r=20$. The solid blue line corresponds 
to the case $\rb\,||\, \bb$, and for the dashed blue line $\rb \perp \bb$.  
The red-dotted line is the input distribution (\ref{eq:GM}) with (\ref{eq:Sb}).   
}   
\label{fig:angle}    
\end{figure}

In Fig.~\ref{fig:angle} we plot $N(r,b,\theta;Y)$ as a function of $b$ for a large dipole with 
$r=20$. Two rapidities were considered.  
In both cases we have compared calculation with $\rb\, ||\, \bb$, solid line, 
and $\rb \perp \bb$, dashed line (together with the input distribution, dotted line).  
The calculation with parallel orientation shows a characteristic peak at $b = \frac{r}{2}$.  
This corresponds to the situation when one end of the dipole, $\xb$, is situated in the center of the target  
where the field is strong $\left<U(\xb)\right>\approx 0$ and the other end, $\yb$, is located in the area where  
the field is very weak $U(\yb)\approx 1$. Thus  the amplitude  
$N=\left<1-U^{\dagger}(\xb) U(\yb)\right> /N_c \simeq 1$.  
On the other hand the large dipole perpendicular to the impact parameter axis has both ends in the region  
where the field is not so strong  and therefore  $N (\xb,\yb)_{\perp}
< N (\xb,\yb)_{||}$.  
 
\subsection{Black disc radius and unitarity bound}  
\label{sec:black}  
  
\begin{figure}[t]  
\centering\includegraphics[width=0.8\textwidth]{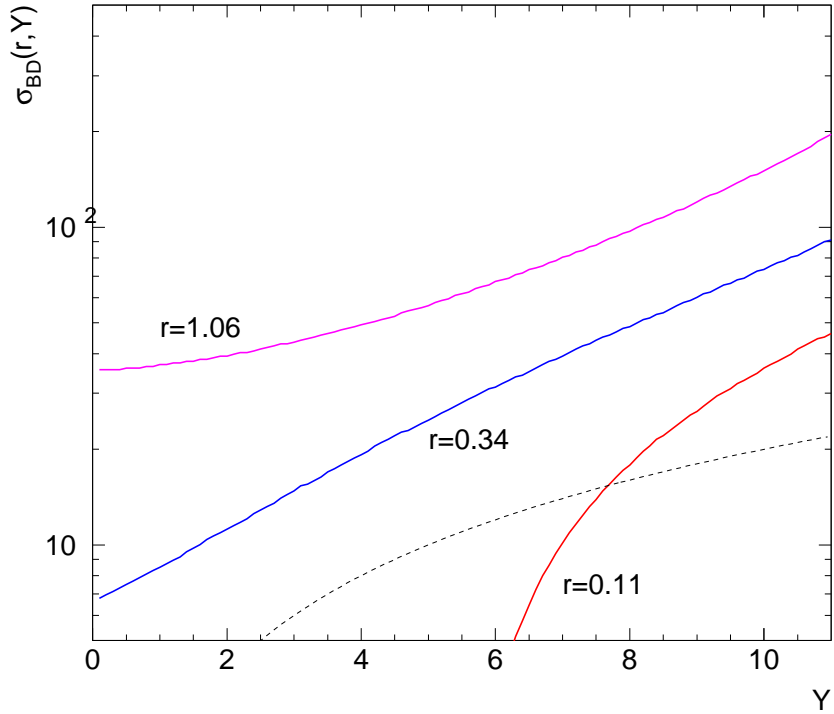}  
\vspace{-1.0cm}  
\caption{The dipole cross section $\sigma_{BD}(r,Y)$ computed from  
Eq.~(\ref{eq:sigmaBD}),  
plotted  as a function  of rapidity $Y$ for three different dipol sizes $r$. 
The dashed line corresponds to $\sigma_{Test}=c\, Y$.}  
\label{fig:bdradius}  
\end{figure}  
  
The black disc radius $R_{BD}(r,Y)$ defines the region in the impact parameter  
space where  
the amplitude $N$ saturates (for a given dipole size $r$ and rapidity $Y$). It is defined  
as the solution of the equation  
\be  
\left<\,N(r,b=R_{BD},\theta,Y)\,\right>_{\theta}\,=\,\kappa  \; ,
\label{eq:rbdcb}
\ee  
with respect to the impact parameter $b$. As before in Sec.~\ref{subsec:satscale}  in our analysis we  
choose $\kappa=1/2$ and average over the angle $\theta$.  
Thus the black disc radius is a function of the dipole size and rapidity.

The expansion of the black disc area with increasing $Y$ is very important  
for the behaviour of the total dipole-nucleus cross section with energy  
$s\sim s_0 e^Y$,  
\be  
\label{eq:4.5}  
\sigma(\rb,Y)\,=\,2\int d^2\bb\, N(\rb,\bb,Y)\,.  
\ee  
This cross section is bounded from below by the cross section integrated  
over the black disc area  
\be  
\label{eq:sigmaBD}  
\sigma_{BD}(\rb,Y) \, = \,2 \int d^2 \bb\, N(\rb,\bb,Y)\,  
\Theta(N-\kappa)\, \simeq\,  
2\,\pi\, R_{BD}^2(r,Y) \, .  
\ee  
Thus the problem of the Froissart bound \cite{FrMa}, $\sigma(\rb,Y)\le Y^2$  
for asymptotically high energies,  can be explicitly studied by looking at  
the rapidity  
dependence of the black disc radius.  
  
In Fig.~\ref{fig:bdradius} we plot $\sigma_{BD}$ from (\ref{eq:sigmaBD})  
as a function of rapidity.  
For the comparison we illustrate the $\sigma_{Test} = c\,Y$ behaviour 
which would be present if the initial profile in impact parameter $S(b)\sim \exp(-b^2)$ was preserved. 
Clearly the behaviour of the black disc cross section is much faster than 
the linear dependence in rapidity. We have found  that 
\be
R_{BD}=R_{BD}^0 \exp(\lambda_{BD}\, \asb Y) \; .
\ee
The extracted value of  $\lambda_{BD}\simeq 0.6 - 0.7$ for $r=0.3-1$ 
and $\lambda_{BD}\simeq 0.9-1$ for $r=3- 30$ at the highest rapidity $Y=11$. 
 
The value of $\lambda_{BD}$ increases rapidly for dipole sizes smaller than the given range above,  
mostly due to the pre-asymptotic effects. The value of $\lambda_{BD}$ is   
smaller than $\lambda_{BD}=0.87\, {\omp}/(2\asb)\simeq 1.2$ which was  quoted in Ref.~\cite{KW}.  
The origin of this discrepancy can be roughly explained by noting  that the power given in \cite{KW}  
most probably overestimates the growth of the black disc radius because it was derived from the analysis  
of the saddle point solution to the linear equation.
 On the other hand in our numerical simulations
the value of $\lambda_{BD} = 0.9-1.0$ for large $r$ is probably closer to the asymptotic one. It is due to the fact that for large $r$, the amplitude has a clear peak for $b=r/2$ which is generated entirely through the evolution.
In other words the amplitude in this region is less sensitive to the
initial profile in $b$. It is also interesting that this value
$\lambda_{BD}$ is consistent with the relation $\lambda_{BD} = 1/2
\lambda_{s}$, based on the following  argument of  the conformal invariance of the
equation (see also \cite{MISHADIS}).

Let us take the saddle point solution to the dipole version of the
BFKL equation in the transverse space \cite{Mueller:95,Salam:96}
\be
N(r_0,r,b,Y) \, \sim \, \frac{r r_0}{16 b^2} \frac{1}{Y^{3/2}} \ln\frac{16
  b^2}{r r_0} \exp\left(\omp Y - \ln^2 \frac{r r_0}{16 b^2}\frac{1}{14
  \zeta(3) \asb Y}\right) \; .
\ee
The amplitude is a function of $r r_0/b^2$ only instead of $r,b$
separately, which is the consequence of the conformal invariance.
The conditions for the saturation scale (\ref{eq:satscb}) and black
disc radius (\ref{eq:rbdcb}) mean
that (roughly)
\be
\frac{r}{R_{BD}^2(r,Y)} \; =\; \frac{1}{b^2 Q_s(b,Y)} \; \longrightarrow
\; \frac{R_{BD}^2(r,Y)}{Q_s(b,Y)} = r b^2 \; .
\label{eq:rbdqsat}
\ee
Since in the last equality the right hand side does not depend on the
rapidity so should  not the left hand side too. This means that if we
have $R_{BD} \sim \exp(\lambda_{BD} \asb Y)$ and $Q_s \sim
\exp(\lambda_s \asb Y)$ then
\be
\lambda_{s} \; = \; 2 \lambda_{BD} \; .
\ee
Relation (\ref{eq:rbdqsat}) means also that $Q_s \sim \frac{1}{b^2}$,
which we have already checked to be approximately true (at least for
large values of $b>7$ see
Sec.~\ref{subsec:satscale}) and also $R_{BD}^2 \sim r$. We have
verified that $R_{BD}^2(r,Y=11) \sim r^a$ with $a \simeq 0.7$ for
$r=0.1 - 1.0$  and $a \simeq 1.0 - 1.1$ for $r=5.0 - 10.0$.

 Clearly the onset of the exponential behaviour  
in rapidity proportional to  $\exp(\lambda_{BD} \asb Y)$, visible in Fig.~\ref{fig:bdradius},  
signals violation of the Froissart bound. This behaviour is a consequence of  
the power tails in impact parameter $N \sim b^{-\gamma}$, in the same way as the steep  
exponential profile $\sim \exp(-b^2)$ leads to the linear dependence $\sim Y$ of the cross section,  
compare the  argument of Heisenberg in \cite{Heisenberg}. The exponential increase of the cross section  
with rapidity despite the multiple scattering interactions has been also observed in  
the Monte Carlo study of the amplitude for onium-onium scattering \cite{Salam:96}.

Since the power-like tail is generated by the long-range contribution, 
the violation of the Froissart bound is caused by the long-range Coulomb-like 
interactions \cite{KW}, which in the reality should be suppressed due to confinement. Thus a modification of the 
BK equation by confinement effects 
is necessary.

\section{Conclusions}

The solution to the BK equation with the $b$-dependence presented in this paper  
differs substantially from the one  with translational invariance.  
The most important result is the fact that the exponential profile in $b$  
in the initial condition is {\em not preserved}  
by the evolution. Instead,  the power behaviour  is generated for large  
$b$'s whose origin comes  
from the structure of  
the BFKL kernel. Such  behaviour leads to the violation of the Froissart  
unitarity bound  
despite the presence of a local unitarity at fixed impact parameter. The  
violation of this bound  
is caused by non-suppressed long range contribution, i.e. the lack of  
confinement in the BK equation.  
 This feature of the $b-$dependent amplitude is  consistent  
 with numerical studies of onium-onium scattering with multiple scattering  
of dipoles \cite{Salam:96},  
 and with the qualitative analysis of the BK equation \cite{KW}.  
  
Thus in order to satisfy the Froissart bound through the evolution 
one would have to modify the evolution kernel in the region of long range  
contribution in order  
to incorporate confinement. For example, it is evident from  
Fig.~\ref{fig:sholo},  
that the naive cut-off -- dropping the second term in the square brackets  
in Eq.~(\ref{eq:ShortLong}) --  
would do the job. In general, conformal symmetry of the integral kernel in  
the BK equation  
has to be broken.  
  
Another interesting feature is the dependence of the solution on the  
dipole  size, which shows  
that the amplitude is saturated in the limited range of scales:  
$1/Q_s(b,Y) < r < R_H(b,Y)$.  
Of course, the true behaviour for large dipole sizes should be modified by  
the long-distance,  
confinement physics.

\section*{Acknowledgments}    
    
We would like to thank    
Jochen Bartels, Edmond Iancu, Alex Kovner, Jan Kwieci\'nski, Misha
Lublinsky, Leszek Motyka,    
Al Mueller, Misha Ryskin, Gavin Salam and Urs Wiedemann for interesting discussions.   
K.G-B is grateful to  Deutsche Forschungsgemeinschaft for a fellowship.    
This research was supported by the Polish Committee for Scientific Research    
grants Nos.\ KBN~2P03B~051~19, 5P03B~144~20.    
    

\end{document}